\documentclass[sigconf]{acmart}
\usepackage{pdfpages}
\pdfoutput=1
\usepackage{booktabs} 
\usepackage{multirow}
\usepackage{subfigure}
\setcopyright{rightsretained}

\acmDOI{}

\acmISBN{123-4567-24-567/08/06}
\acmConference{The Web Conference}{May 2019}{San Francisco, CA, USA}
\acmYear{2019}
\copyrightyear{2019}

\acmArticle{4}
\acmPrice{15.00}

\begin{document}
\title[Neural IR Meets Graph Embedding]{Neural IR Meets Graph Embedding: \\ A Ranking Model for Product Search}

\author{Yuan Zhang}
\affiliation{
  \department{Key Laboratory of Machine Perception (MOE)}
  \department{Department of Machine Intelligence}
  \institution{Peking University, China}
}
\email{yuan.z@pku.edu.cn}

\author{Dong Wang}
\affiliation{
  \department{Center for Data Science}
   \institution{Peking University, China}
}
\email{wangdongcis@pku.edu.cn}

\author{Yan Zhang}
\affiliation{
  \department{Key Laboratory of Machine Perception (MOE)}
  \department{Department of Machine Intelligence}
  \institution{Peking University, China}
}
\email{zhy@cis.pku.edu.cn }

\begin{abstract}
Recently, neural models for information retrieval are becoming increasingly popular. They provide effective approaches for product search due to their competitive advantages in semantic matching. However, it is challenging to use graph-based features, though proved very useful in IR literature, in these neural approaches. 
In this paper, we leverage the recent advances in graph embedding techniques to enable neural retrieval models to exploit graph-structured data for automatic feature extraction. The proposed approach can not only help to overcome the long-tail problem of click-through data, but also incorporate external heterogeneous information to improve search results. Extensive experiments on a real-world e-commerce dataset demonstrate significant improvement achieved by our proposed approach over multiple strong baselines both as an individual retrieval model and as a feature used in learning-to-rank frameworks.
\end{abstract}

%
%



\maketitle

\section{Introduction}
\label{intro}
Online shopping has become increasingly popular over the past decade. Those e-commerce sites normally sell a vast number of products at the same time and provide plenty of choices under each category. For instance, there are hundreds of million products in Amazon and over one billion in eBay. It is almost impossible for consumers to find items that they are interested in without the help of an effective product search engine. On the other side, an improvement in product search accuracy could bring massive revenue to online retailers \cite{ai2017learning}.

In a typical product search scenario, users formulate queries using characteristics of the product of interest, for example, the producer's name, brand, or terms which describe the category of the product \cite{rowley2000product}. 
However, the lexical gap between queries and product descriptions \cite{van2016learning} makes it difficult to exploit such traditional information retrieval models as LSI \cite{deerwester1990indexing}, BM25 \cite{Robertson:2009:PRF:1704809.1704810}, which focus more on lexical matching, in this application scenario. 

The recent neural network-based models (sometimes also known as deep learning models) \cite{mitra2017neural} provide a good solution to this problem with their advantages in semantic matching. 
Yet, despite great success of deep learning in processing speech (1D), image (2D), and video (3D) data, neural models are rather limited in dealing with \textit{high-dimensional} graph-structured data, which, on the other hand, has been extensively studied in the existing IR literature (e.g., the graph-based ranking features PageRank \cite{page1999pagerank}, HITS \cite{kleinberg1999authoritative}; the graph-based similarity metric SimRank \cite{jeh2002simrank}; the click graph-based smoothing and propagation models \cite{craswell2007random, gao2009smoothing,li2008learning, wu2013learning,jiang2016learning}; the heterogeneous graph-based query intent learning \cite{ren2014heterogeneous}).

\begin{figure}[t] 
\centering
\includegraphics[width=8cm]{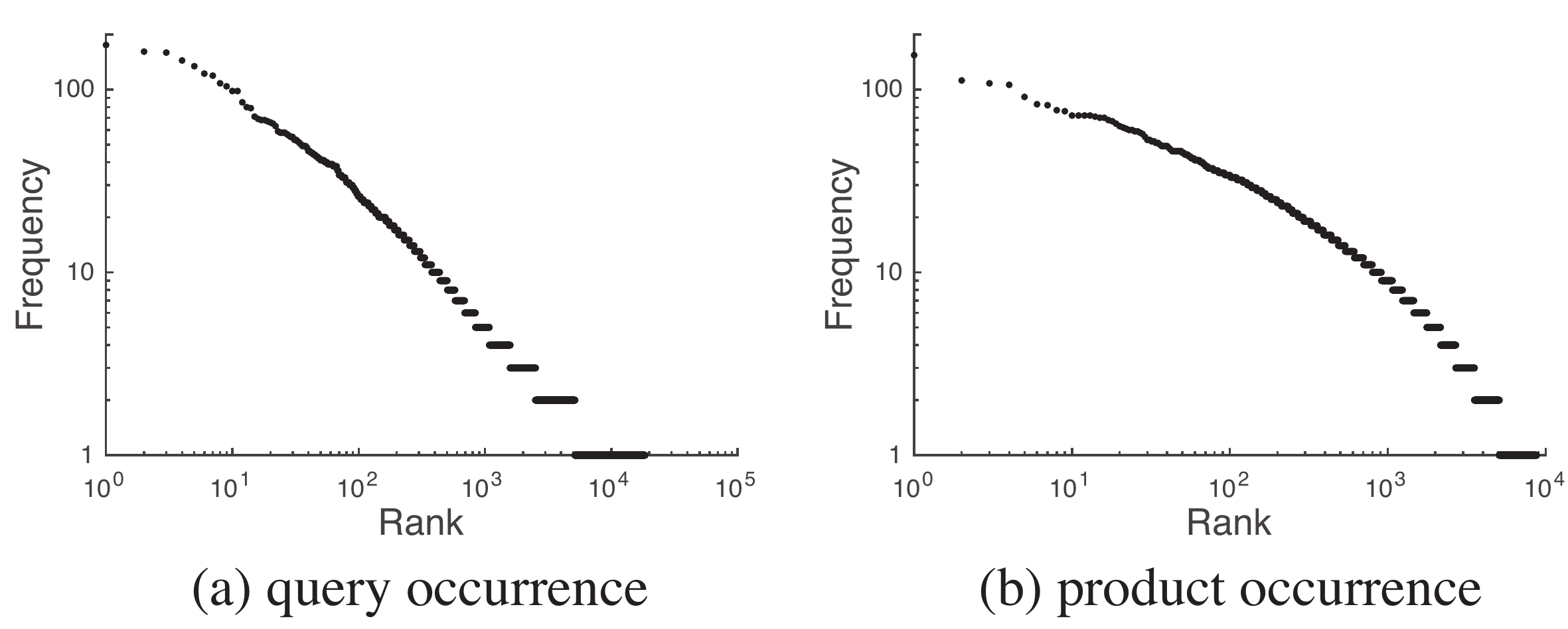} 
\caption{The frequency-rank distribution (log-scale) of item and query occurrences in a sample of search logs from an e-commerce search engine, respectively. 
}
\label{fig:longtail}
\end{figure}

\begin{figure}[t] 
\centering
\includegraphics[width=5.8cm]{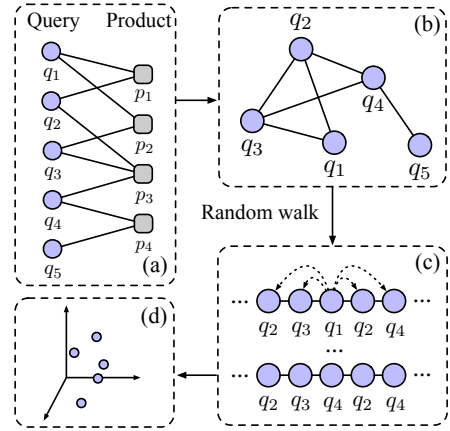} 
\caption{Graph embedding.}
\label{fig:background}
\end{figure}

In fact, the neural approach and the graph-based approach are based on two different types of \textit{inductive biases} that are complementary to each other. The inductive bias for the former is the smoothness with respect to input features (i.e., $f(x + \epsilon) \approx f(x)$ for small $\epsilon$). For example, given queries ``multifunction printer'' and ``all-in-one printer'', neural nets should give similar predictions. On the contrary, the graph-based approach is mostly based on the smoothness regarding node proximities where node can be queries, items, sessions, etc. (i.e., $f(x) \approx f(y)$ if node x and y are ``close'' in a given graph). For example, queries $q_1$ and $q_4$ sharing common neighbors within a densely connected subgraph in Figure \ref{fig:background} are assumed to have similar outcomes and therefore the knowledge towards one query can be transferred to the other. Admittedly, in theory, neural nets can automatically learn the latter structure with sufficiently large amount of data. However, besides extensive computational burden, it is infeasible to obtain enough data for long-tailed queries/items in e-commerce search (Figure \ref{fig:longtail}) and extremely hard for neural nets to learn high-order proximities in practice. Thus, it becomes an interesting research question whether we can combine these two complementary approaches for better retrieval performance.

In this paper, we propose to leverage the recent advances in \textit{graph} \footnote{In order to avoid confusion, we mostly use the word \textit{graph} to refer to the collection of nodes and edges, and \textit{network} to refer to the neural networks in this paper, although these two words are often used interchangeably in literature.} \textit{embedding} techniques \cite{Goyal2017graph,hamilton2017representation} to integrate graph-structured data into a unified neural ranking framework. 
To the best of our knowledge, this is the first study on how to use the click-graph features in neural models for retrieval.

The benefits to allow neural retrieval models to incorporate graph-structured information are demonstrated in two-fold. 
Firstly, by integrating the click-graph features, our model can effectively overcome the query-item sparsity problem of the click-through data \cite{gao2009smoothing} and generalize better to unseen queries and long-tailed products.
This is of great significance in the deep learning setting, where neural models not only require a large amount of data with sufficient coverage but also are prone to overfitting, which in turn causes poor retrieval performance on rare queries and reinforces the \textit{rich-get-richer} effect \cite{Cho2014impact}.
Furthermore, we show that graphs can be used as convenient and efficient tools in our approach to fuse external heterogeneous information to improve search results, e.g., co-viewed items in a browsing session and co-purchased items in an order. Those features are, though, not very straightforward for existing neural ranking models to take advantage of.

From another perspective, our proposed approach can be seen as a way to tackle a side effect of neural networks, \textit{unawareness of transitivity}, as their capacity (which can be loosely measured by the number of parameters) grows to fit non-smooth relations with limited data. 
For instance, as illustrated in Figure \ref{fig:background}(a), query $q_4$ and $q_5$ share a common product click so that they tend to be queries with similar user intent and product $p_3$ (clicked through $q_4$) should also be considered somewhat relevant to $q_5$.
However, neural network models themselves might not be aware of this transitive relation, because query-product pair $(q_5, p_3)$ simply do not appear in the positive training examples and, when given sufficient capacity, neural networks do not require this relation to hold in order to preserve other positive relations either (see Section \ref{sect:discussion} for empirical evidence).
While this issue can be implicitly mitigated to some extent with such regularization methods as L1/L2 penalties that merely ``uninformatively'' shrink the weights towards zero or with carefully designed neural architectures,
it strongly motivates the direct use of graph-structured regularization towards utilizing the full potentials of neural networks for retrieval.

Finally, we evaluate our model on a real-world dataset of e-commerce search. Experimental results show that our model substantially outperforms state-of-the-art baselines both as an individual ranking model and as a feature in a learning-to-rank framework. Ablation study further demonstrates that the use of graph-embedding techniques leads to a significant improvement in retrieval performance. 

To conclude, the major contributions of this work are as follows:

\begin{itemize}
	\item We propose a Graph Embedding-based ranking model for Product Search (GEPS), which is, to the best of our knowledge, the first to integrate click-graph features into a unified neural ranking framework. 
	\item Our proposed graph embedding-based approach can effectively (a) deal with the long-tail sparsity problem and (b) combine heterogeneous external information into neural models to improve search results.
	\item Through extensive experimental study on a publicly available e-commerce dataset, we show that GEPS can achieve significant improvement in retrieval performance compared with the existing work.
\end{itemize}

\begin{figure*}[t] 
\centering
\includegraphics[width=16cm]{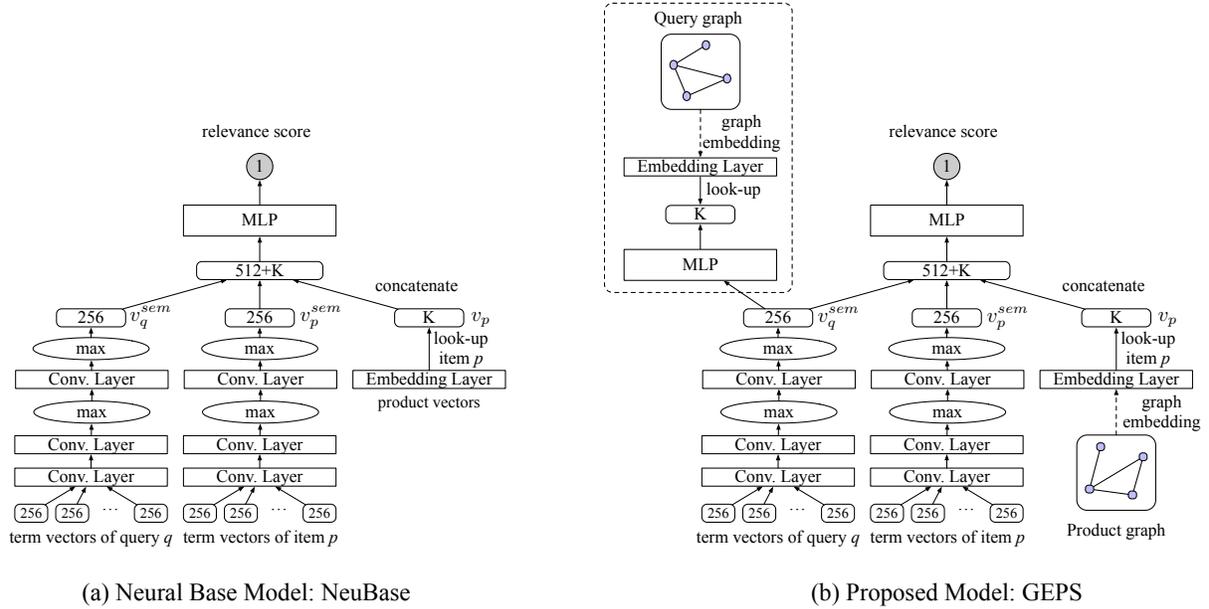} 
\caption{Model Architecture. We use dashed-line arrows to represent assignment of graph embedding vectors to the parameters/weights in embedding layers. $K$ denotes the embedding dimension.}
\label{fig:architecture}
\end{figure*}

\section{Background: Graph Embedding}
As mentioned in Section \ref{intro}, there are many ``hand-crafted'' graph-based features in the existing IR literature. However, the lack of an effective representation makes it hard to apply deep learning models to graph-structured data for automatic feature extraction. Recently, the extensive study on the \textit{graph embedding} provides a good solution to this problem. In this section, we introduce two popular state-of-the-art graph embedding methods, Deepwalk \cite{perozzi2014deepwalk} and LINE \cite{tang2015line}, by taking the query graph induced by click-through data as an example.

As shown in Figure \ref{fig:background}(a-b), we construct a query graph by establishing links between each pair of queries if they have once led to clicks on an identical item. (A product graph can also be constructed similarly.) 
This query graph contains rich signals for retrieval, for edges in the graph may indicate semantic similarities of queries. Intuitively, two queries that are nearby in the query graph tend to be similar to each other, and hence the knowledge we have for one query can also be transferred to the other, which can be especially useful for a tail query that rarely appears in search logs.
Motivated by this kind of intuition, the goal of graph embedding is to obtain vector representations for each node in a graph, as shown in Figure \ref{fig:background}(d), so that the proximities between nodes can be well preserved. 

Perozzi et al. \cite{perozzi2014deepwalk} propose Deepwalk based on a distributional hypothesis on graphs, nodes with similar neighborhood tend to be similar to each other, which is also called the \textit{second-order proximity}. They first generate random walk sequences and then use the node in the middle to predict the other nodes within a given window (Figure \ref{fig:background}(c)), so that nodes occurring in similar contexts of random walk sequences will have close embeddings, as in Skip-gram \cite{mikolov2013distributed} developed in language modeling. Thus, query $q_1$ and $q_4$ are supposed to have similar vector representations.

Further, Tang et al. \cite{tang2015line} propose LINE to take the \textit{first order proximity} into consideration, for example, query $q_4$ and $q_5$ are connected by a direct link and therefore should have close embeddings as well. Meanwhile, graph embedding is made scalable to graphs with over millions of nodes and billions of edges by an edge sampling technique proposed in their work. Finally, the embedding vectors preserving the two kinds of proximities are obtained respectively in LINE, and they can either be used separately or combined in a supervised manner.

\section{Model Framework}
\label{sect:model}
\subsection{Preliminaries and Notations}
Like other neural ranking models, we use click-through data to train our models. The click-through data used in our study consists of queries and products with their co-clicked information.
We denote the set of queries as $Q$ and each query is represented by a collection of terms $q = \{t^{(q)}_1, ..., t^{(q)}_M\}$.
Similarly, the set of products is denoted by $P$ and each product by $p=\{t^{(p)}_1, ..., t^{(p)}_N\}$. We also denote as $w$ the word vector for each term $t$ with the same subscripts and superscripts, for example, $w^{(q)}_1$ denotes the word vector of $t^{(q)}_1$.
For each query $q \in Q$, the set of clicked items are represented by $P_q$, and items that are presented in the SERP (search engine results page) but not clicked as $\overline{P_q}$. 

The query graph and the product graph are defined as $G_{query} = (Q, E_{Q})$ and $G_{product} = (P, E_{P})$, respectively. After graph embedding, we can obtain graph embedding vectors $e_q$ for each query $q \in Q$ and $e_p$ for each product $p \in P$. The graph embedding procedures then can be defined as mappings, $g:G(V,E) \mapsto \{e_v, \forall v \in V\}$.

While our proposed model framework is not limited to any specific graphs $G_{query}$ and $G_{product}$ as long as they can provide meaningful interaction information between their nodes (i.e., queries and products) for search, the reader can assume that the graphs of queries and products are induced by the click-through data as depicted in Figure \ref{fig:background}(a-b) until further discussion in Section \ref{sect:fuse}.

\subsection{Proposed Model: GEPS}
Our goal is to combine the advantages of both graph-based methods and neural approaches. In this section, we first introduce a simple yet typical neural network architecture for product search as the ``base model'', and then show how graph embedding techniques can be plugged into such neural network models to incorporate graph-structured information for better retrieval performance.

As shown in Figure \ref{fig:architecture}(a), we represent terms of queries and product descriptions by their term vectors pre-trained on the whole corpus. Then, we input these term vectors to convolutional layers for semantic feature extraction, where the filter window is set to three. Max pooling layers are used to reduce feature dimensions and eventually transform feature maps into a feature vector. We also use an id embedding vector $v_{p}$ for each product, which is commonly used in e-commerce search engines to capture its intrinsic features not reflected in the product description. The semantic feature vector $v_{q}^{sem}$ for the query and $v_{p}^{sem}$ for the product along with the product vector $v_p$ are concatenated into a long vector. Finally, we use an MLP (multi-layer perceptron) to output the relevance score.

To integrate graph-based features for products, we use graph embeddings to take the place of product id embeddings $v_p$ in the network architecture described above. This is analogous to using pre-trained word embeddings instead of learning word vectors from scratch, which has already been proved to be both effective and efficient in natural language processing. We find it rather helpful to fine-tune these graph embeddings during training due to the unsupervised nature of graph embedding methods at hand.

While it is straightforward to incorporate product graph embeddings in the neural framework, making use of the query graph embeddings is quite challenging, because our retrieval model inevitably has to handle lots of new queries that are absent in the click-through data for training and hence do not have corresponding graph embedding vectors.
\footnote{Although there could also exist products missing in the click-through data, the proportion is much smaller in practice. We initialize them with all-zero vectors meaning without any prior knowledge. Empirically, this trick works well enough.} 
A tempting approach would be training a regression model $f$ (e.g., CNN or RNN) with query terms as input that ``mimics'' the query graph embedding vectors, that is,
\begin{equation}
arg\min_f  \Vert f(t^{(q)}_1, ..., t^{(q)}_M) - e_q \Vert_2^2,
\end{equation}
and then use $f$ to re-create embedding vectors for queries \footnote{A similar approach \cite{D17-1010} is used by Pinter et al. to obtain embedding vectors for the out-of-vocabulary (OOV) words with a character-level RNN.}.  
However, we find it hard to fine-tune query graph embeddings with this approach to achieve best performance; fine-tuning either the re-created vectors themselves or the model $f$ could easily cause overfitting. Besides, it is not ideal to consider graph-based features and semantic features separately to model textual queries (different from the item side).

In our model, we propose to use an encoder-based approach to encode query-graph features in their corresponding semantic feature vectors. More specifically, we use an MLP to transform the semantic feature vectors $v_{q}^{sem}$ into the same vector space as query graph embeddings $e_{q}$ and try to minimize the reconstruction error as shown in Figure \ref{fig:architecture}(b), that is,
\begin{equation}
\label{eq:re}
\begin{aligned}
min. & \quad \mathcal L_{re}(q) = \Vert \widetilde{e_{q}} - e_{q} \Vert_2^2 \\
st. &\quad \widetilde{e_{q}} =  MLP(v_{q}^{sem}).
\end{aligned}
\end{equation}
Different from the other approach mentioned above, we can jointly optimize this reconstruction loss with the final ranking loss in the model training stage (see Section \ref{sect:training}).
Also, this encoder-based approach has an additional effect: the word vectors are further trained to capture relevance-based semantics \cite{zamani2017relevance} more accurately in such a way that the semantic vectors $v_{q}^{sem} = CNN(w^{(q)}_1, ..., w^{(q)}_M)$ can better reconstruct high-order relevance relations encoded in the query graph. This approach actually shares a similar idea to \textit{multi-task learning}, where relevant tasks (analogous to the graph embedding reconstruction here) are trained simultaneously for mutual regularization and knowledge transfer.

Finally, it is worth mentioning that the proposed graph embedding-based approach can be used as a ``plug-in'' to boost the performance of almost any other neural retrieval model. Thus, the experimental results in this paper merely provide a lower bound for the performance of our proposed approach since we only use a simple base model for ease of demonstration.

\subsection{Fusing Heterogeneous Information}
\label{sect:fuse}

\begin{figure}[t] 
\centering
\includegraphics[width=5.9cm]{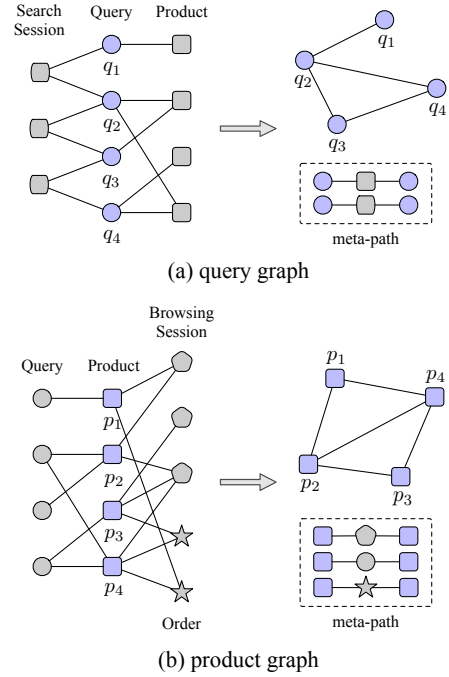} 
\caption{Fusing heterogeneous information with meta-paths.}
\label{fig:heter}
\end{figure}
Our proposed approach can utilize rich information on interactions between both queries and products contained in \textit{graphs}, so that the knowledge we obtain from training data can be transferred to similar queries and correlated products for more accurate retrieval results. 
To fully take advantage of our graph-based approach, we can further incorporate fruitful external heterogeneous information, in addition to click-through data, into the graphs we use. 

As shown in Figure \ref{fig:heter}(a), a search session may contain several queries; we can assume that the queries within a search session are more likely to share similar search intent. By regarding the heterogeneous relations among sessions, queries and products as a heterogeneous graph \cite{ren2014heterogeneous}, we can use meta-paths \cite{Sun11pathsim:meta} to construct the query graph.
The semantics of graphs constructed by meta-paths are well interpretable, for example, the edge set $E_{query}$ in Figure \ref{fig:heter}(a) contains all pairs of queries that are either issued in a search session or lead to clicks of an identical product. 

We can also fuse more heterogeneous relations in the product graph as in Figure \ref{fig:heter}(b). 
In an e-commerce website, in addition to direct search, customers also browse products through product category hierarchies, recommendations, etc.; products that are browsed at the same time can also be considered as somewhat related to each other. This can become especially useful in the cold-start scenarios where the products that are never clicked through any search queries. Besides, co-purchased items in the same order provide useful signals for search as well. Similarly, we make use of meta-paths to construct the product graph $G_{product}$.

Admittedly, more irrelevant factors and even data noises (e.g., unrelated queries in a session) are introduced into graphs as we fuse more heterogeneous information. 
Fortunately, our model is trained in a supervised manner so that the neural nets only extract useful features in terms of product search to produce the final results. Again, our proposed approach combines the advantages of both graph-based methods and deep learning approaches.

\subsection{Model Training}
\label{sect:training}
In the training stage, we use a pairwise ranking loss to train our neural model. Given a search query $q$ along with a positive product $p_+ \in P_q$ and a negative product $p_- \in \overline{P_q}$, the loss is defined as
\begin{equation}
\mathcal L_{rank}(q, p_+, p_-) = h(score(q, p_+) - score(q,p_-)),
\end{equation}
where $score(q, p)$ is the relevance score given by our model for query-item pair $(q, p)$ and $h$ is the \textit{smoothed hinge loss} \cite{Rennie05lossfunctions},
\begin{equation}
h(x) = 
\begin{cases}
    0   &  x \geq 1, \\
    1/2 - x &  x < 0,\\
    (1 - x)^2 /2  &  \text{otherwise}.
 \end{cases} 
\end{equation}
At the same time, we also minimize the reconstruction error as given in Eq. (\ref{eq:re}) for encoding graph embedding-based features in the query semantic vectors. To sum up, the overall loss function is
\begin{equation}
\label{eq:loss}
\mathcal L = \sum_{q \in Q}\sum_{p_{+} \in P_q}\sum_{p_{-} \in \overline{P_q}} \mathcal L_{rank}(q, p_+, p_-)  + \alpha \cdot \sum_{q \in Q} \mathcal L_{re}(q),
\end{equation}
where $\alpha$ is a tunable parameter (in our experiments, we just empirically set $\alpha = 1$).

We apply the stochastic gradient descent (SGD) method with mini-batches to optimize the loss function given in Eq. (\ref{eq:loss}). Meanwhile, the standard weight decay regularization and early stopping strategy are used to prevent overfitting. We also use gradient clipping in case gradient explosion happens. The batch size is set to 1024 and the weight decay is $10^{-4}$ in our experiments. 


\section{Experiments}
\label{sect:experiments}
\subsection{Research Questions}
\label{research_questions}
We are particularly interested in whether and, if so, how our proposed graph embedding-based approach can lead to performance gains in product search. We break down this principal research question into four individual ones to guide our experimental study: 

\begin{itemize}
\item[\textbf{RQ1.}] How does our proposed model GEPS compare to the state-of-the-art retrieval models?
\item[\textbf{RQ2.}] How does our graph embedding-based approach benefit neural models in dealing with the long-tail problem of click-through data?
\item[\textbf{RQ3.}] How does our graph embedding-based approach further benefit neural models by incorporating external heterogeneous information to improve search results? 
\item[\textbf{RQ4.}] How do graph embedding algorithms influence the performance of GEPS?
\end{itemize}

\subsection{Experimental Setup}
\subsubsection{Dataset and Data Partition} 
In order to answer those research questions in Section \ref{research_questions}, we evaluate our proposed model on the dataset from CIKM Cup 2016 Track 2 \footnote{https://competitions.codalab.org/competitions/11161}, which is also the only publicly available dataset of product search. The dataset contains five months of logs of an e-commerce search engine from 1st Jan to 1st Jun, 2016. We filter out the queries without textual information, for the search engine also allows ``query-less'' search.

In the \textbf{original data split}, the test set is obtained by taking the last queries of some sessions. 
However, we have to keep in mind that, with this data partitioning procedure, there could be larger overlaps between training topics and test topics than in real application scenarios, both lexically and semantically, for mainly two reasons: the queries in the same session but prior to a test query are very likely to share similar search intent; both the training and test sets are evenly distributed in the five-month period, not being significantly influenced by the topic shift in queries over time.

\begin{table}[t]
\centering
\caption{Dataset statistics.}
\label{dataset}
\begin{tabular}{@{}ccc@{}}
\toprule
Statistics                     & \multicolumn{2}{c}{Value}                                                              \\ \midrule
\# products             & \multicolumn{2}{c}{184047}                                                             \\
\# unique  queries & \multicolumn{2}{c}{26137} \\
Vocabulary size                & \multicolumn{2}{c}{181194}                                                             \\
Length of queries              & \multicolumn{2}{c}{2.66 $\pm$ 1.77}                                                     \\
Length of product descriptions  & \multicolumn{2}{c}{5.12 $\pm$ 2.04}                                                      \\
\# items per order      & \multicolumn{2}{c}{1.34 $\pm$ 0.96}                                                      \\ \midrule
                               & \multicolumn{1}{c}{Original} & \multicolumn{1}{c}{Chronological} \\ \cmidrule(l){2-3} 
\# queries (train/test) & 35615/16218                             & 28380/3011                                   \\
\# search sessions      & 26880                                   & 21505                                        \\
\# browsing sessions    & 349607                                  & 242852                                       \\
\# clicks          & 37705                                   & 30160                                        \\
\# views           & 1235380                                 & 857008                                       \\
\# orders               & 13506                                   & 9130                                         \\ \bottomrule
\end{tabular}
\end{table}

In order to simulate the real-world application scenario where models are trained using historical data and applied to unseen future queries, we also report experimental results with a \textbf{chronological data split}. We use the first four months' logs (until 1st May) for training and use the last month's test data for evaluation.
Note that, in this data split, we still cannot know those hidden search behaviors in the first four months (because they are not provided in the dataset), so the size of training data is only about 80 percent of that in the original data split and hence results on those two testbeds are not reasonably comparable. Also, the statistical significance of performance improvement could subsequently decline as the size of test data is reduced to 20 percent.

The detailed statistics about the dataset and data partitioning are listed in Table \ref{dataset}.

\subsubsection{Baselines}
We conduct comparison study with the following baseline methods.
\begin{itemize}
\item \textbf{LSI} (Latent Semantic Indexing) \cite{deerwester1990indexing} is a classical vector space model in information retrieval. We set the dimension of vector spaces $K = 256$.

\item\textbf{BM25} \cite{Robertson:2009:PRF:1704809.1704810} is a very popular probabilistic model-based ranking function proposed by Robertson et al.

\item \textbf{QLM} (Query-likelihood Model) with Jelinek-Mercer smoothing \cite{zhai2004study} is an effective entity search approach based on language models \cite{van2016learning}. 
For each product $p$, we construct a statistical language model smoothed by the whole corpus,
\begin{equation}
P_\lambda(t | p) =(1 - \lambda) \cdot P_{ml}(t | p) + \lambda \cdot P(t | C),
\end{equation}
where $P_{ml}(t|p)$ and $P(t|C)$ are simply given by relative counts of term $t$ appearing in the description of product $p$ and the whole corpus $C$, respectively. Then, we can obtain the relevance score for each product $p$ given a query $q$ according to Bayes' rule,
\begin{equation}
P(p | q) \propto P(p) \cdot P_\lambda(q|p) = P(p) \cdot \prod_{t \in C_q} P_\lambda(t|p),
\end{equation}
where $P(p)$ is the popularity of product $p$ given by its click counts, and $C_q$ is the collection of terms in query $q$.
In our experiments, we set the smoothing parameter $\lambda = 0.01$ that turns out to be not very sensitive.

\item \textbf{LSE} (Latent Semantic Entities) \cite{van2016learning, VanGysel2017sert} is a state-of-the-art latent vector space model for product search, which jointly learns vector representations of words and products from scratch. We choose 256 as the dimensions of both word vectors and product vectors.

\item \textbf{VPCG} (Vector Propagation on Click Graphs) \cite{jiang2016learning} is a state-of-the-art non-neural graph-based retrieval model that utilizes vector propagation on click graphs to overcome the above-mentioned long-tail problem and the lexical gap between queries and documents.

\item \textbf{ARC-II} \cite{hu2014convolutional}, \textbf{MatchPyramid} \cite{pang2016text} are two state-of-the-art neural retrieval models . We use the implementations from the open-source toolkit MatchZoo\footnote{https://github.com/NTMC-Community/MatchZoo}\cite{fan2017matchzoo} specifically designed for neural IR. We set the embedding size to 256 and report the results achieved by the best of multiple model setups provided in MatchZoo.

\item \textbf{NeuBase} (Figure \ref{fig:architecture}(a)) is a variant of GEPS that does not use graph-based features (i.e., excluding the graph embedding-based components from GEPS) so as to investigate the performance improvement achieved by the proposed approach. In fact, this model is an extension of \textbf{C-DSSM} \cite{shen2014learning}.
\end{itemize}

Since there exists little overlap between terms of queries and product descriptions (about 1.8\%) \cite{wu2017ensemble} in our dataset, traditional retrieval models (LSI, BM25 and QLM) and LSE (originally based on product reviews instead of descriptions) from our baseline pools exhibit poor performances.
For a more rigorous comparison, we also report the results of these models with augmented product descriptions by adding the terms of queries that led to clicks to the corresponding products. 

\subsubsection{Evaluation Metrics}
In our evaluation, we regard the items returned by the e-commerce search engine in the SERP for each test query as candidates, and evaluate ranking performance of different models on those candidates. The products shown in the SERP for each test query are automatically labeled based on user-specific actions by three grades of relevance, \textit{irrelevant} (0), \textit{somewhat relevant} (1), \textit{very relevant} (2), which correspond to \textit{not clicked}, \textit{clicked}, and \textit{clicked and purchased} products, respectively.

Three standard evaluation metrics, MRR (Mean Reciprocal Rank), MAP (Mean average precision) and NDCG (Normalized Discounted Cumulative Gain), are used for comparison. MRR measures the rank position of the first relevant (i.e., either \textit{somewhat relevant} or \textit{very relevant}) product which is important to attract users to keep browsing, while MAP measures the rank of all relevant products. In addition to the metrics that assume binary relevance, NDCG is used to capture multiple relevance levels, where product purchase is considered as the ultimate goal of the e-commerce search task.

\subsection{As an Individual Retrieval Model}

\begin{table}
\centering
\caption{Retrieval performance. $\dagger$ indicates statistically significant improvement ($p < .01$) by the pairwise t-test over all the other baselines. We report performance gains in NDCG relatively to the original search results returned by the commercial search engine without re-ranking. 
}
\label{tab:performance}
\resizebox{8.5cm}{!}{
\begin{tabular}{llcccc} \toprule
\multicolumn{2}{c}{Models}                                                                                             & MRR                       & MAP                       & NDCG                      & Gain (\%)           \\ \midrule
\multicolumn{6}{r}{\textbf{Original data split} }                                                                                                                                                                                \\
\multicolumn{1}{c}{}                                                                                    & LSI          & .2223                     & .2117                     & .3536                     & +3.07               \\
                                                                                                        & BM25         & .2250                     & .2143                     & .3545                     & +3.33               \\
                                                                                                        & QLM          & .2422                     & .2305                     & .3670                     & +6.99               \\
                                                                                                        & LSE          & .2313                     & .2162                     & .3577                     & +4.26               \\
\multicolumn{1}{c|}{\multirow{4}{*}{\begin{tabular}[c]{@{}c@{}}w/ \\ augmented\\ documents \end{tabular}}} & LSI          & .3787                     & .3575                     & .4618                     & +34.63              \\
\multicolumn{1}{c|}{}                                                                                   & BM25         & .2897                     & .2652                     & .3869                     & +12.77              \\
\multicolumn{1}{c|}{}                                                                                   & QLM          & .4743                     & .4449                     & .5173                     & +50.80              \\
\multicolumn{1}{c|}{}                                                                                   & LSE          & .4357                     & .4075                     & .4958                     & +44.52              \\
                                                                                                        & VPCG         & .4331                     & .4091                     & .4928                     & +43.65              \\
                                                                                                        & ARC-II       & .4128                     & .3849                     & .4776                     & +39.23            \\
                                                                                                        & MatchPyramid & .4792                     & .4498                     & .5228                     & +52.40              \\
                                                                                                        & NeuBase      & .4448                     & .4162                     & .4982                     & +45.22              \\
                                                                                                        & GEPS         & $\mathbf{.4927}^\dagger$  & $\mathbf{.4677}^\dagger$  & $\mathbf{.5368}^\dagger$  & $\mathbf{+56.49}$   \\ \hline
\multicolumn{6}{r}{ \textbf{Chronological data split} }                                                                                                                                                                          \\
                                                                                                        & LSI          & .2119                     & .2021                     & .3467                     & +1.37               \\
                                                                                                        & BM25         & .2219                     & .2122                     & .3543                     & +3.62               \\
                                                                                                        & QLM          & .2271                     & .2175                   & .3587                    & +4.89               \\
                                                                                                        & LSE          & .2185                     & .2054                     & .3509                     & +2.62               \\
\multicolumn{1}{c|}{\multirow{4}{*}{\begin{tabular}[c]{@{}c@{}}w/ \\augmented\\ documents \end{tabular}}} & LSI          & .3129                     & .2949                     & .4182                     & +22.30              \\
\multicolumn{1}{c|}{}                                                                                   & BM25         & .2632                     & .2450                     & .3790                     & +10.84              \\
\multicolumn{1}{c|}{}                                                                                   & QLM          & .3603                     & .3370                     & .4487                     & +31.22              \\
\multicolumn{1}{c|}{}                                                                                   & LSE          & .3542                     & .3310                     & .4445                     & +29.98              \\
                                                                                                        & VPCG         & .3571                     & .3391                     & .4505                     & +31.72              \\
                                                                                                        & ARC-II       & .3388                     & .3165                     & .4350                     & +27.20              \\
\multicolumn{1}{c}{}                                                                                    & MatchPyramid & .3864$^\dagger$                   & .3631$^\dagger$                     & .4692 $^\dagger$                    & +37.21              \\
                                                                                                        & NeuBase      & .3475                     & .3263                     & .4444                     & +29.96              \\
                                                                                                        & GEPS         & $\mathbf{.3907}^\dagger$  & $\mathbf{.3684}^\dagger$  & $\mathbf{.4753}^\dagger$  & $\mathbf{+39.00}$   \\ \bottomrule
\end{tabular}}
\end{table}

\subsubsection{Retrieval Performance}
\label{sect:retrieval}
To answer \textbf{RQ1} we list the results of baselines and the proposed model GEPS on the real-world testbed in Table \ref{tab:performance}. 
As mentioned in Section \ref{intro}, the traditional lexical-based models, LSI and BM25, do not perform well in the product search scenario even with augmented product descriptions by clicked queries. 
In contrast, neural retrieval models (i.e., ARC-II, MatchPyramid, NeuBase and GEPS) exhibit strong performances due to their advantages in semantic matching.
Meanwhile, the click-graph based unsupervised model VPCG also provides a competitive baseline especially in the chronologically split dataset, which demonstrates the effectiveness of graph-based features in product search.
By incorporating graph-structured data into the supervised neural retrieval model with the proposed graph embedding-based approach, GEPS can substantially outperform all the baseline methods in both data split settings, which proves the ability of GEPS to combine the advantages of graph-based models and neural network models.

\begin{figure}[t] 
\includegraphics[width=7cm]{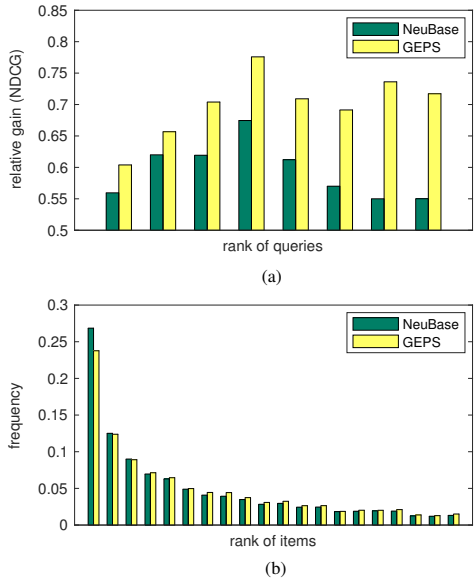} 
\caption{Retrieval performance regarding the long-tail problem of click-through data. (a) Relative gains in NDCG w.r.t. the popularity ranks of queries. (b) Frequencies of products appearing at the top-5 positions of ranking results returned by NeuBase and GEPS w.r.t. their popularity ranks.}
\label{fig:tail_exp}
\end{figure}

To answer \textbf{RQ2}, we divide search queries into eight groups according to their occurrence frequency ranks in the click-through data as in Figure \ref{fig:longtail}(a) and plot the averaged relative performance gains of NeuBase and GEPS in each group in Figure \ref{fig:tail_exp}(a) \footnote{Due to limited space, we only report results in the original data split when conducting model analysis in the rest of this paper. Similar results can be obtained in the chronological data split.}. In the last four groups where increasingly fewer training samples are provided, the performance of NeuBase starts to decline dramatically, while GEPS still remains robust. 
(Both of their relative gains increase in the first half of groups because the commercial search engine performs relatively better on popular queries.) 
This suggests that neural retrieval models indeed suffer from overfitting problems on those rare queries in the training data and our proposed graph embedding-based approach provides an effective solution. 

We also report the frequencies of products that appear at the top-k ($k = 5$) positions of the ranking results obtained by NeuBase and GEPS with regard to their frequencies of occurrence in the training data. Figure \ref{fig:tail_exp}(b) shows that NeuBase tends to rank popular products higher and tail products lower than GEPS, which not only hurts search accuracy but also reinforces the \textit{rich-get-richer} phenomenon \cite{Cho2014impact} that might cause some long-term issues for e-commerce sales platforms. This further demonstrates that the proposed graph embedding-based approach can enable neural models to exploit graph-structured relations to make smoother predictions on the long-tailed products.

\begin{table}[t]
\centering
\caption{Ablation study in the original data split. \textit{P\_emb} and \textit{Q\_emb} denotes the graph embeddings of products and queries respectively, while \textit{Heter} represents the integration of external heterogeneous information into models. We use $+$/$-$ to denote cases with/without a certain feature in the model. All the improvements relative to their previous rows are statistically significant ($p < .01$) unless otherwise denoted by an asterisk.}
\label{tab:ablation}
\begin{tabular}{@{}ccccccc@{}}
\toprule
\multicolumn{1}{l}{P\_emb} & \multicolumn{1}{l}{Q\_emb} & \multicolumn{1}{l}{Heter} & MRR & MAP & NDCG & Gain (\%) \\ \midrule
- & - & - &.4448&.4162&.4982&+45.22 \\
+ & - & - &.4529&.4275*&.5076&+47.97 \\
+ & - & + &.4640&.4387&.5160&+50.43 \\
+ & + & - &.4810&.4540&.5256&+53.22  \\
+ & + & + &.4955&.4704&.5378&+56.76 \\ \bottomrule
\end{tabular}
\end{table}

\subsubsection{Ablation Study}
\label{sect:ablation}
To investigate \textbf{RQ3} and further answer \textbf{RQ2}, we conduct an ablation study of our proposed model. The results are shown in Table \ref{tab:ablation}. We can conclude that incorporating graph embeddings of both products and queries can lead to significant performance improvement as a result of its successful treatment towards the long-tail problem of click-through data by comparing row $2$ and row $4$ to the baseline result row $1$.

In addition, we can observe substantial performance gains when comparing row $2$ and row $4$ with their extended versions with external heterogeneous information in row $3$ and row $5$, respectively. This shows that the proposed approach is effective in taking advantage of heterogeneous information to improve search accuracy.

\begin{figure}[t] 
\centering
\includegraphics[width=6.5cm]{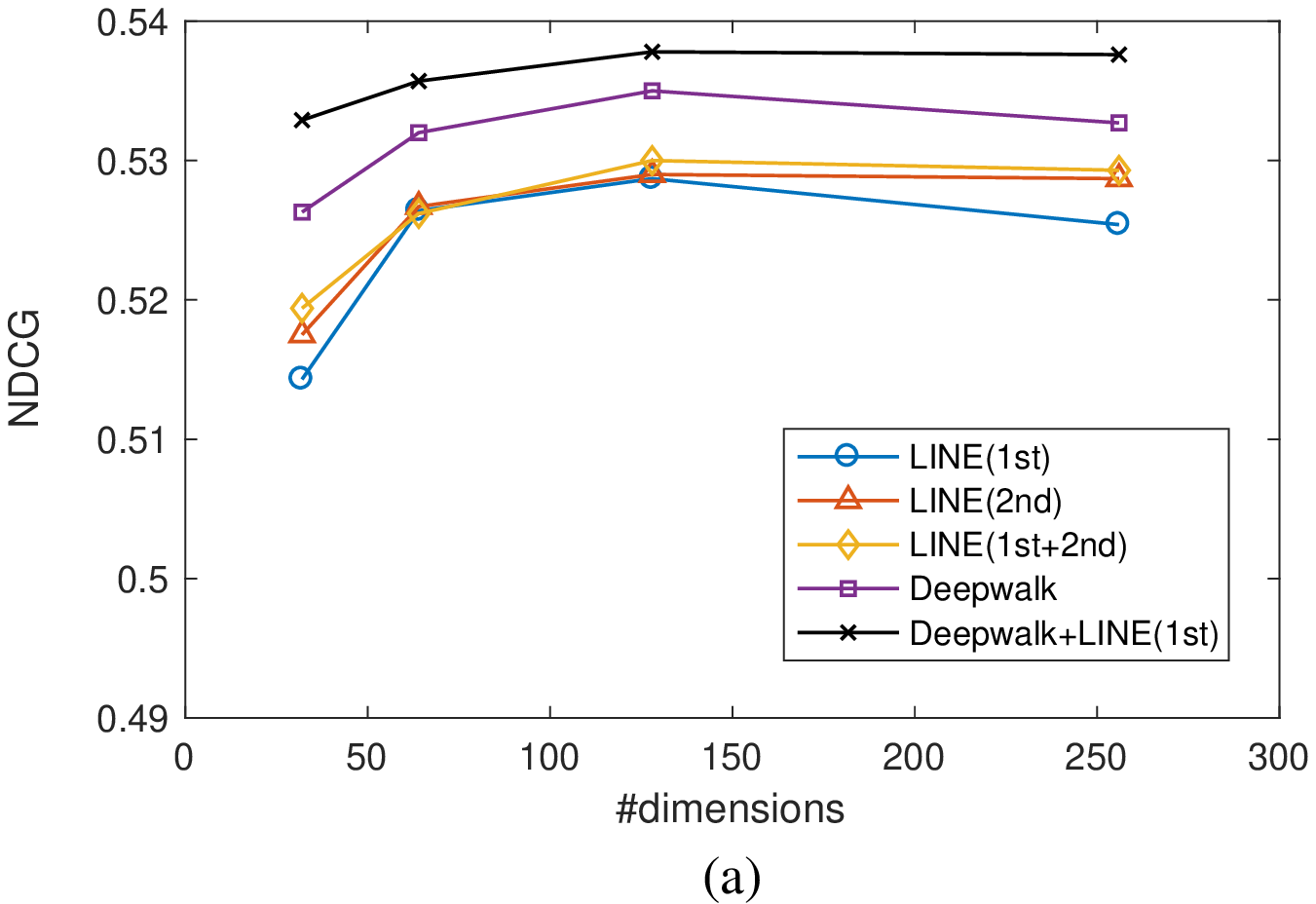} 
\includegraphics[width=6.5cm]{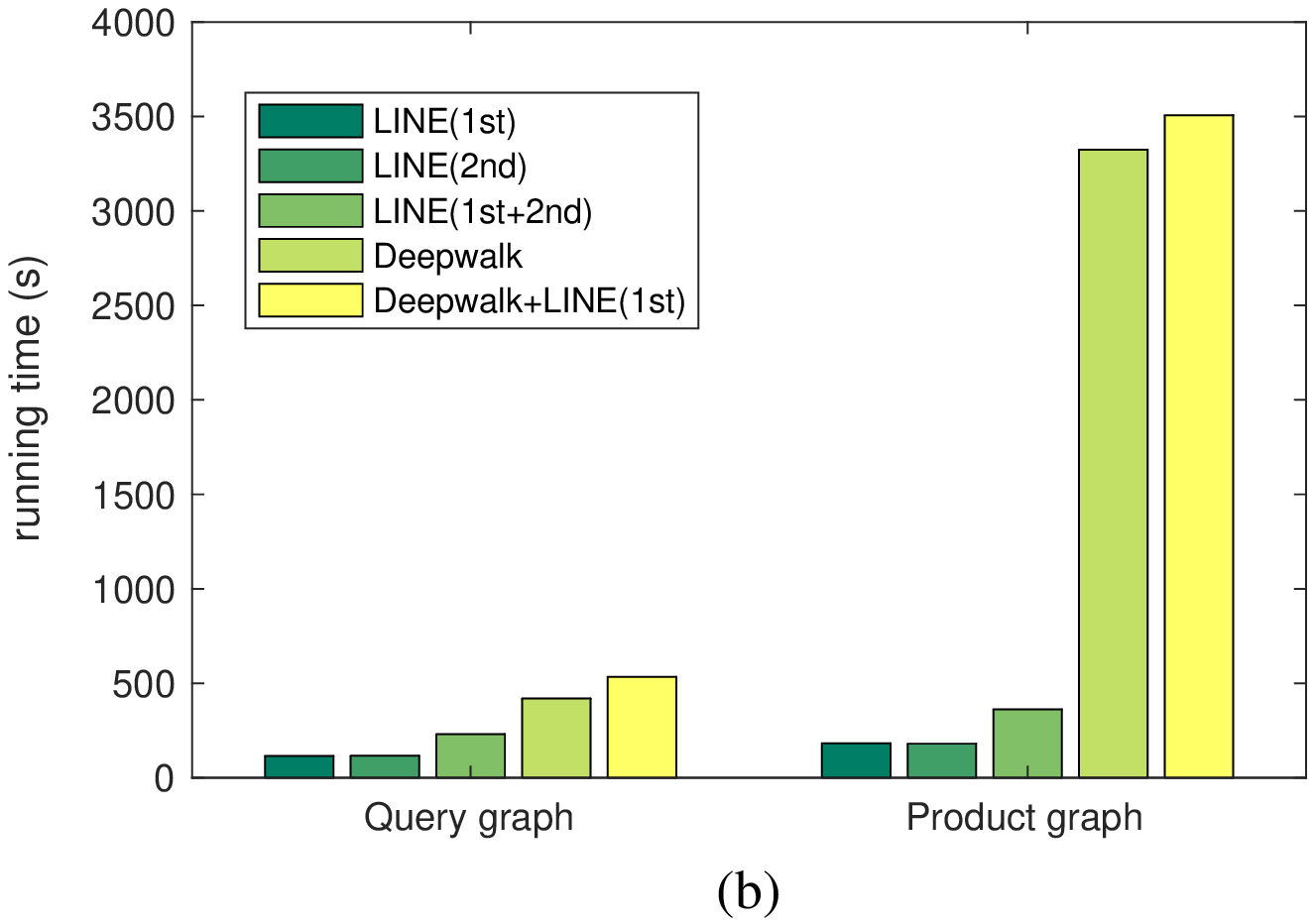}
\caption{Performance of the graph embedding methods w.r.t. different number of dimensions $K\in [32, 64, 128, 256]$.}
\label{fig:para}
\end{figure}
\subsubsection{Analysis of Graph Embedding Methods}
\label{sect:net_emb}
To answer \textbf{RQ4}, we analyze the performance of different graph embedding methods. As shown in Figure \ref{fig:para}(a), Deepwalk performs slightly better than LINE with both embedding vectors preserving the first order proximity, the second order proximity, and their combination. 
This difference can be seen as a trade-off between efficacy and efficiency, for LINE runs relatively faster than Deepwalk as shown in Figure \ref{fig:para}(b) thanks to its edge-sampling strategy (although they both have linear time complexity).
Besides, it can lead to another marginal gain to concatenate embedding vectors obtained from Deepwalk with the \textit{first order} embedding vectors obtained from LINE, because Deepwalk only considers the \textit{second order} proximity to which the latter are complementary. 

We also investigate how embedding dimensions influence the final performance. Overall, our proposed approach is quite robust to different graph embedding methods with different embedding sizes (Figure \ref{fig:para}(a)). Throughout this paper, we use ``Deepwalk + LINE (1st)'' with $K=128$ as the default setting.

\subsection{As a Feature for Learning-to-rank Models}
\label{sect: l2r}

\begin{figure}[t]
\centering
\includegraphics[width=7cm]{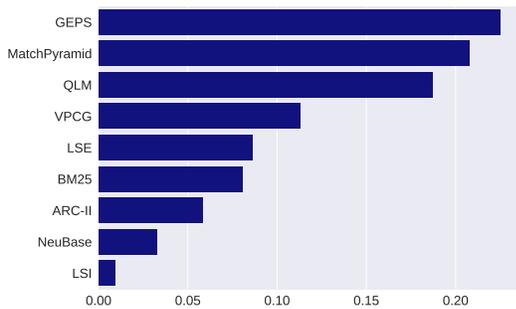} 
\caption{Importance scores of features used in GBDT.}
\label{fig:ltr}
\end{figure}

\begin{table}[t]
\centering
\caption{Retrieval performance as a feature in the learning-to-rank framework. 
}
\label{tab:ltr_results}
\resizebox{8.5cm}{!}{
\begin{tabular}{@{}lcccc@{}}
\toprule
\multicolumn{1}{c}{Models} & MRR & MAP & NDCG & Gain (\%) \\ \midrule
\multicolumn{5}{r}{\textbf{Original data split}} \\
Base & .4249	&.3973	&.4876	&+42.14  \\ \midrule
Base+QLM & .4878	&.4568	&.5266	&+53.52 \\ 
Base+LSE & .4450	&.4160	&.5000	&+45.75 \\
Base+VPCG & .4414	&.4178	&.4992	&+45.50 \\
Base+MatchPyramid&.4880	&.4579	&.5291	&+54.24\\
Base+GEPS & $\mathbf{.4991}$	&$\mathbf{.4717}$	&$\mathbf{.5381}$	&$\mathbf{+56.87}$ \\ \midrule
Base+QLM+LSE & .4828	&.4525	&.5227	&+52.36 \\
Base+LSE+MatchPyramid&.4976	&.4673	&.5359	&+56.22\\
Base+QLM+MatchPyramid & .5113 &	.4791 & .5434 & +58.40\\
Base+LSE+GEPS & .5021	&.4744	&.5402	&+57.46\\
Base+MatchPyramid+GEPS & .5120	& .4827	& .5459	&+59.14\\
Base+QLM+GEPS & $\mathbf{.5207}$	&$\mathbf{.4913}$	&$\mathbf{.5520}$  & $\mathbf{+60.91}$ \\ 
\midrule
Base+QLM+MatchPyramid+VPCG	&.5124	&.4819	&.5454	&+58.97\\
Base+QLM+LSE+MatchPyramid	&.5162	&.4843	&.5475	&+59.60\\
Base+QLM+LSE+GEPS&.5218	&.4919	&.5524	&+61.03\\
Base+QLM+MatchPyramid+GEPS	&$\mathbf{.5289}$	&$\mathbf{.4986}$	&$\mathbf{.5573}$	&$\mathbf{+62.45}$\\
\bottomrule
\end{tabular}}
\end{table}

We continue to answer \textbf{RQ1} by evaluating the performance of GEPS as a feature in a learning-to-rank framework implemented with GBDT \cite{friedman2001greedy}. We choose seven handcrafted features motivated by Wu et al. \cite{wu2017ensemble} as our basic feature pool \textit{Base}: (0) original ranks (which might contain some unreleased information used by the commercial search engine), (1) total click counts, (2) total view counts, (3) total purchase counts, (4) click-through rates, (5) the word lengths of the description of each product, (6) the logarithms of product price. 

We can draw the following conclusions from the results shown in Table \ref{tab:ltr_results}. First, GEPS is such a powerful feature that we can observe the largest performance improvement by adding GEPS to the feature pool. Second, GEPS is also complementary to other competitive retrieval approaches, especially the traditional language-modeling based approach \textit{QLM} and the recently proposed neural architecture \textit{MatchPyramid}; combining GEPS with QLM and MatchPyramid leads to a substantial improvement by 3.6\% in NDCG over the best single model.
However, it is interesting that we do not observe a significant improvement after combining GEPS with those features in the basic feature pool, possibly because most of the useful basic features can be automatically extracted from click-through data by GEPS.
In addition, in Figure \ref{fig:ltr}, we can see that GEPS is the most important feature when ensembling all the baselines with GBDT.

\section{Further Discussion: Retrieval as Link Prediction and Transitivity}
\label{sect:discussion}

\begin{figure}[t] 
\centering
\includegraphics[width=7cm]{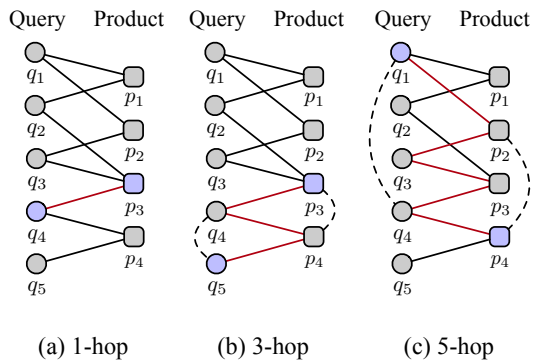} 
\caption{(a) Retrieval problems can be regarded as link prediction. (b-c) Products within several hops (e.g., 3 and 5 hops away) to a certain query tend to be relevant as well. In GEPS, such transitive relations are regularized by similar graph embeddings indicated by the dashed lines.}
\label{fig:trans}
\end{figure}

The goal of product search, especially from the perspective of neural network-based black box approaches, is to predict whether a given query-product pair $(p, q)$ is relevant (and then rank all products according to the probability). In fact, this goal is exactly equivalent to predict whether a \textit{link} exists in the bipartite click graph of queries and products as shown in Figure \ref{fig:trans}(a). 

Nonetheless, the click graph used in the training stage can never be complete; there are inevitably many missing links partly because we can only access a limited sample of search logs and also because, given a query, some of the relevant products are not presented to customers by the search engine at all (though, this does not mean those products have no user impressions within other queries either). The good part is that we can obtain more information beyond those direct links thanks to the \textit{transitivity} of relevance relations. To be more specific, products that are several hops from a given query (e.g., $q_5$ versus $p_3$ in Figure \ref{fig:trans}(b) and $q_1$ versus $p_4$ in Figure \ref{fig:trans}(c)) are more likely to be relevant than a totally random product. However, neural network models are not well aware of such transitivity although the one-hop relations can be perfectly maintained due to their strong expressive power.

In our proposed approach, we can effectively capture transitivity with graph embedding techniques, where correlations between queries and products are well preserved.  For instance, in Figure \ref{fig:trans}(b), $q_4$ and $q_5$ as well as $p_3$ and $p_4$ have similar graph embeddings (i.e., $e_{q_4} \approx e_{q_5}$ and $e_{p_3} \approx e_{q_4}$) since they are close neighbors (i.e., the \textit{first order proximity}) in the query graph and the product graph respectively, and therefore $p_3$ is considered as a relevant item to query $q_5$ because both features $[e_{q_4}, e_{p_3}]$ and $[e_{q_5}, e_{p_4}]$ indicate positive links in the click graph. It is the same case with 5-hop pair $(q_1, p_4)$ in Figure \ref{fig:trans}(c) where $q_1$ and $q_4$ as well as $p_2$ and $p_4$ have similar neighbors (i.e., the \textit{second order proximity}) and hence are also close to each other in the graph embedding space.

\begin{figure*}[t] 
\centering
\subfigure[NeuBase]{
\includegraphics[height= 3.42cm]{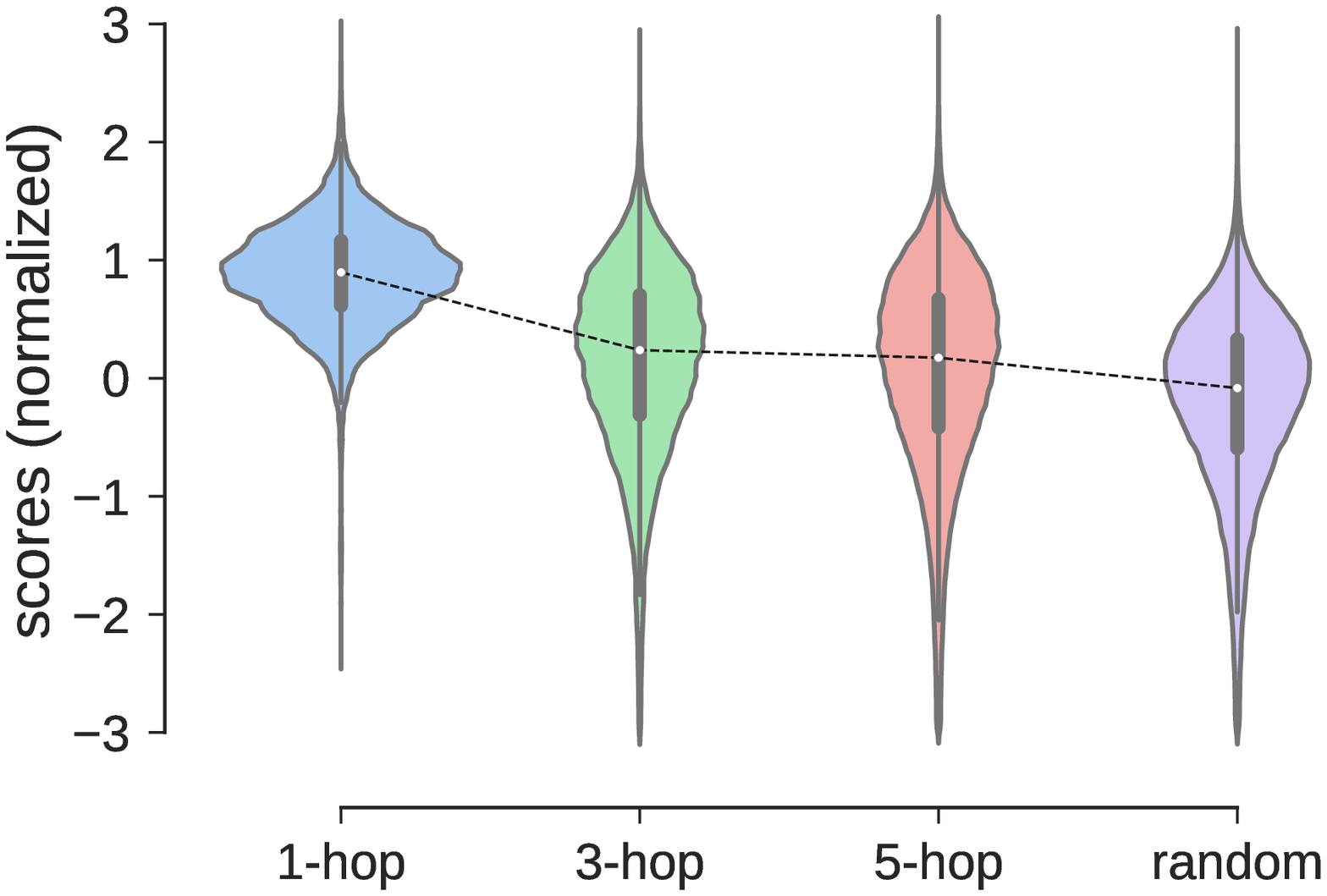} 
}
\subfigure[GEPS]{
\includegraphics[height= 3.42cm]{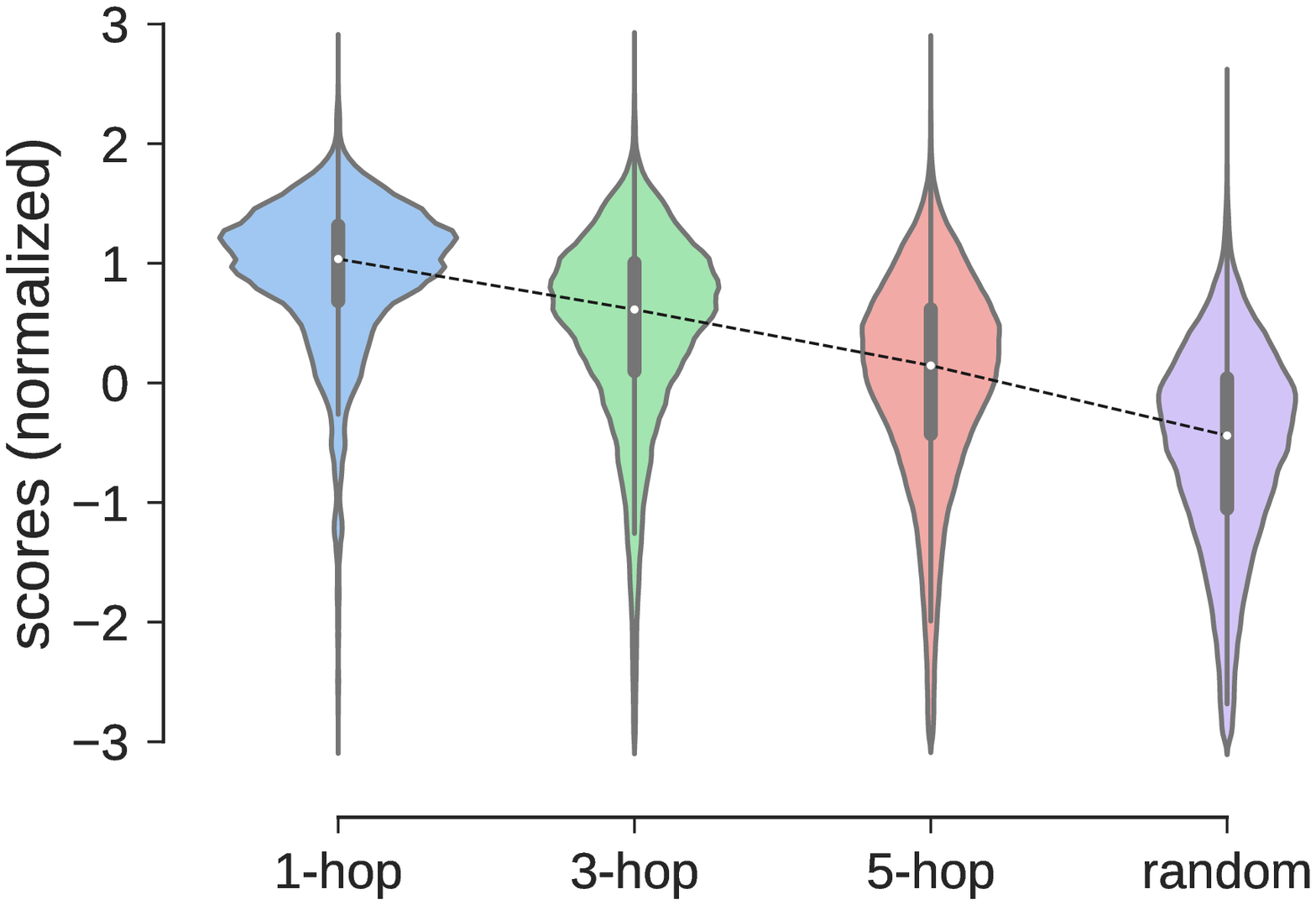} 
}
\subfigure[MatchPyramid]{
\includegraphics[height= 3.74cm]{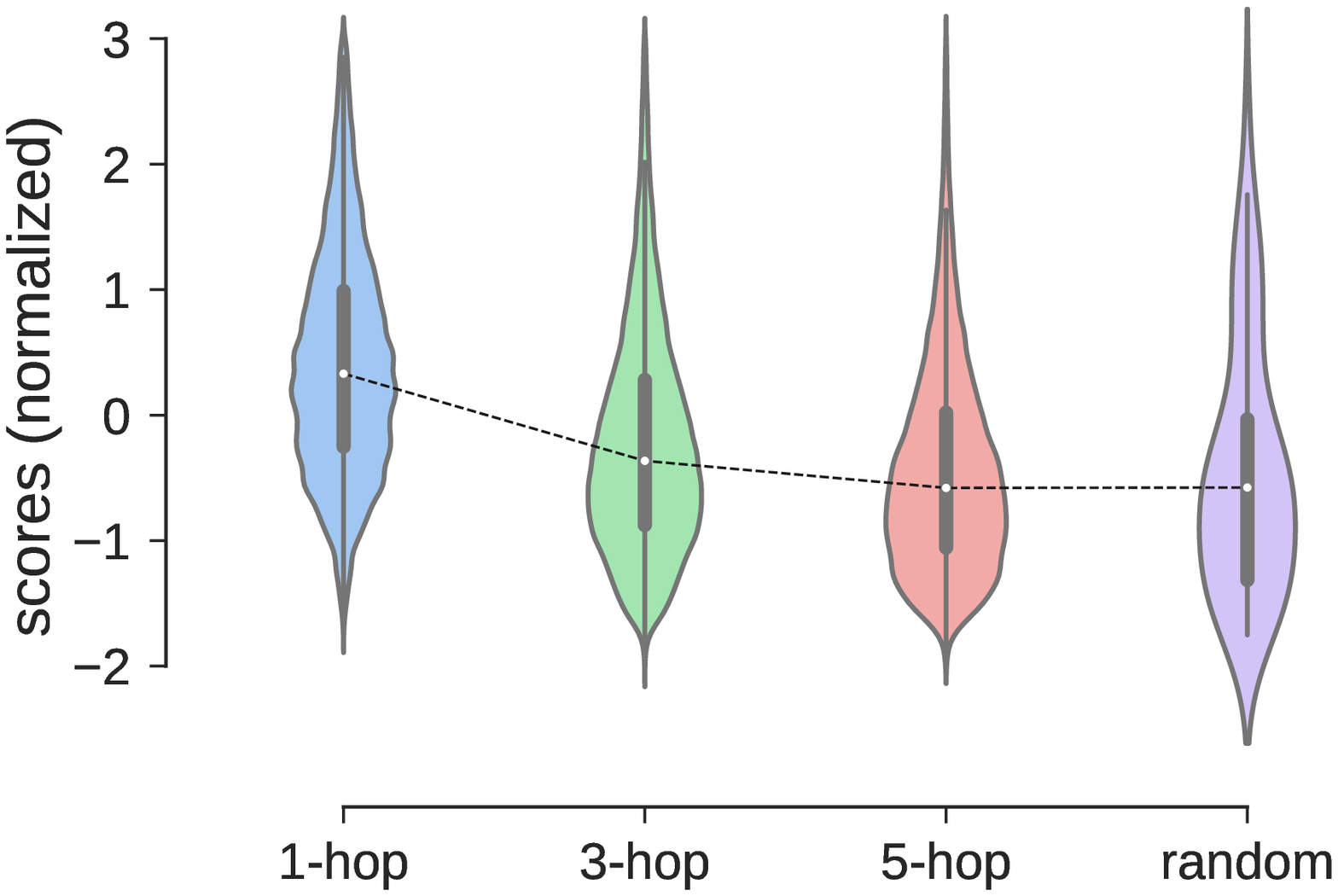} 
}

\caption{The distributions of relevance scores given to the pairs of queries and products that are $k$ hops way from each other and given to random query-product pairs. }
\label{fig:violin}
\end{figure*}

To empirically prove this point, we plot the distributions of relevance scores given by NeuBase, the proposed model GEPS and MatchPyramid for the pairs of queries and products that are \textit{1 hop}, \textit{3 hops} and \textit{5 hops} away in the click graph along with \textit{random} pairs in Figure \ref{fig:violin}. In the case of NeuBase, we can observe that the relevance scores for 3-hop and 5-hop query-product pairs drop quickly and their distributions are very similar to that of random pairs. On the contrary, GEPS can give smoothly decreasing scores as queries and products are farther away from each other in the click graph; the relevance score distribution for 3-hop query-product pairs is closer to that for 1-hop pairs, while the distribution for 5-hop pairs is somewhere between that for 3-hop pairs and that for random pairs. Table \ref{tab:auc} supports our argument in a more rigorous way; without such proper regularization as graph structures, neural networks are likely to fit non-smooth relations without transitivity (e.g., 3-hop and 5-hop cases) as exactly appeared in the training data (i.e., 1-hop query-product pairs). To conclude, it is the ability to discriminate among those decreasingly relevant query-product pairs (3-hop, 5-hop and so on relatively to totally random) induced by transitivity that enables better ranking performance for GEPS. 

Also shown in Figure \ref{fig:violin} and Table \ref{tab:auc}, instead of stratifying query-product pairs of different distances in the click graph, MatchPyramid discriminates positive samples (i.e., 1-hop query-product pairs) in a more fine-grained manner. This suggests that MatchPyramid tackles the issues of neural networks from a different perspective from graph-based approaches (actually by focusing on the semantic interaction between query terms and document terms with a properly designed neural architecture). This is also consistent with the fact that MatchPyramid and GEPS can substantially boost each other's performance as features in the learning-to-rank model as shown in Table \ref{tab:ltr_results} and Figure \ref{fig:ltr} (Section \ref{sect: l2r}).


Furthermore, such transitive relevance relations can also be induced from more complicated graphs than click graphs, for example, the ones shown in Figure \ref{fig:heter}. This is also why it can lead to significant improvement when incorporating external heterogeneous information with our graph-embedding approach.
\begin{table}[t]
\centering
\caption{AUC results of separating the pairs of queries and products within $k$ hop(s) ($k \in [1,3,5]$ ) from farther pairs with relevance scores given by NeuBase and GEPS respectively .}
\label{tab:auc}
\begin{tabular}{@{}lccc@{}}
\toprule
Model   & within 1 hop & within 3 hops & within 5 hops \\ \midrule
NeuBase & .833         & .616          & .627         \\
GEPS    & .824         & .744          & .749          \\ 
MatchPyramid&.726&.664&.647\\\bottomrule
\end{tabular}
\end{table}


\section{Related Work}
We review the related work of product search, neural models for IR and graph-based models for IR along with a brief comment on neural models with graph-structured data in a more general context.
\subsection{Product Search}
Product search has been attracting extensive research efforts. Duan et al. (2013) \cite{Duan:2013:PMM:2541176.2505578} propose a probabilistic generative model to analyze e-commerce search log for useful knowledge. Duan et al. (2015) \cite{Duan:2015:MCI:2806416.2806557} study the problem of learning query intent representation for product search and recommendation with a proposed language model. Karmaker Santu et al. \cite{KarmakerSantu:2017:ALR:3077136.3080838} discuss issues and strategies when applying learning-to-rank models to the product search scenario. Also, Li et al. \cite{Li:2011:TTM:1963405.1963453} propose an economic theory-based product search model to complement those product search models mainly adapted from the information retrieval literature.

Recently, Van Gysel et al. \cite{van2016learning} propose a latent vector space model (LSE) to jointly learn distributional representations of words and products for better product retrieval performance. Ai et al. \cite{ai2017learning} exploit a hierarchical embedding model (HEM) that jointly learns semantic representations for various entities, including terms, products, users and queries, to tackle the problem of personalized search. These two embedding models largely inspire our work although neither of them leverages graph embedding techniques to encode rich graph-structured relationships in their embeddings. 

\subsection{Neural Information Retrieval}
\label{sect:related_neural}
Our work is in the context of recent advances in neural information retrieval. Due to limited space, only some of the recent work is introduced here, and we refer the reader to the paper \cite{mitra2017neural} by Mitra and Craswell for a comprehensive review. 
Huang et al. \cite{huang2013learning} propose a deep structured semantic model (DSSM) to learn semantic representations of queries and documents with large-scale click-through data for web search, while Guo et al. \cite{guo2016deep} propose an interaction-focused deep relevance matching model (DRMM) that uses histogram-based features to capture lexical relevance between queries and documents for better retrieval performance. Mitra et al. \cite{mitra2017learning} then attempt to combine those two approaches with a hybrid deep model Duet composed by a local model for exact term matching and a distributed model for semantic matching. 
Further inspired by the human judgment process, DeepRank \cite{pang2017deeprank} uses deep neural networks to capture local relevance in various relevant locations extracted from a given document and then aggregate such local relevance information to output the global relevance score.
Recently, Liu et al. \cite{P18-1223} introduces knowledge graphs to neural search systems, which, however, is different from the intrinsic problem of transitive relevance in click-through data we study in this paper. 

In fact, our contribution is orthogonal to this line of research and therefore can be combined with the above mentioned neural retrieval approaches.

\subsection{Graph-based Information Retrieval}
Graph-based features are widely used and studied in the IR literature \cite{Mota2011Graph}, e.g., Pagerank \cite{page1999pagerank}, HITS \cite{kleinberg1999authoritative} and Simrank \cite{jeh2002simrank}. Yet, our work is more related to the following. 
Craswell and Szummer \cite{craswell2007random} use random walks on the click graph to produce probabilistic rankings for image search. Gao et al. \cite{gao2009smoothing} propose a random walk-based smoothing approach to overcome the sparsity problem of click-through data in web search. Click graphs are also used to learn semantic representations of queries and documents (e.g., \cite{wu2013learning,jiang2016learning}) and to learn query intent (e.g., \cite{li2008learning, ma2008learning}). Ren et al. \cite{ren2014heterogeneous} propose to use a heterogeneous graph that consists of queries, web pages, and Wikipedia concepts to improve search performance. 

Graph-based IR approaches can depend on global information to make local decisions and hence are rather robust to sparse, complex and incomplete data. However, most of them also suffer from disadvantages as unsupervised methods. While combining them into a learning-to-rank framework as features is one possible way to overcome this limitation, it also motivates us to directly tackle this problem at the model level to combine both the advantages of unsupervised graph-based models and supervised neural models.

\subsection{Neural Models with Graph-structured Data}

In a more general context, GCN (Graph Convolutional Network) \cite{NIPS2016_6081,Kipf2017Semi} provides an alternative solution to incorporate graph-structured data into neural networks by generalizing the traditional convolutional filters from images to graphs.
Berg et al. recently propose a model called GC-MC (Graph Convolutional Matrix Completion) that attempts to apply GCN to recommender systems in an inspiring unpublished preprint \cite{Berg2017Graph}. Monti et al. \cite{monti2017geometric} later extend GC-MC to multi-graph cases and obtain promising results. 

The proposed approach and GCN can both be seen as ways to regularize neural models with graph structures. The former uses an auxiliary task of reconstructing graph embeddings to achieve such goal, whereas GCN directly conditions the output of neural network on the adjacency matrix (i.e., $f(\cdot|A)$). Due to the latter fact, each prediction depends on the whole graph
(or, at least, a local subgraph as approximation \cite{ying2018graph}) 
causing GCN not very scalable to such large graphs as in real-world search scenarios, and even worse, unavoidable absent queries in the training data cannot be properly dealt with. On the contrary, graph embeddings can serve as an ``intermediate'' which decouples neural nets and graphs \textit{per se} to reduce computational burden while still allowing neural nets to utilize graph structure information.

\section{Conclusion}
This paper introduces a Graph Embedding-based ranking approach for Product Search (GEPS) which leverages recent advances in graph embedding techniques to incorporate graph-structured data into neural models in Section \ref{sect:model}. Through extensive experiments in Section \ref{sect:experiments}, we demonstrate that our proposed approach can successfully combine both the advantages of graph-based retrieval models and neural retrieval models. 
To be more specific, experimental results show that GEPS enables neural models to effectively deal with the long-tail sparsity problem and combine heterogeneous external information to improve search accuracy. 
Finally, we point out the equivalence between retrieval tasks and link prediction tasks, and then discuss the advantage of GEPS in capturing relevance transitive relations on graphs in Section \ref{sect:discussion}. 

\bibliographystyle{ACM-Reference-Format}
\bibliography{SIGIR18-bibliography}

\end{document}